\documentclass[lettersize,journal]{IEEEtran}

\ifCLASSOPTIONcompsoc
  \usepackage[nocompress]{cite}
\else
  \usepackage{cite}
\fi

\usepackage[numbers]{natbib}

\usepackage{url}
\usepackage{CJKutf8}
\usepackage{algorithm}
\usepackage{subfigure}
\usepackage{xcolor}
\usepackage{booktabs} 
\usepackage{mathtools}
\usepackage{amsmath,bm}
\usepackage{makecell}
\usepackage{multirow}
\usepackage{multicol}
\usepackage{array}
\usepackage{comment}
\usepackage{amsfonts,amssymb}
\usepackage{textcomp}
\usepackage{fmtcount}
\usepackage{flushend}
\usepackage[skip=3pt]{caption}

\usepackage[noend]{algpseudocode}
\usepackage{algorithmicx,algorithm}

\usepackage{amsmath}
\usepackage{graphicx}
\usepackage{multirow}
\usepackage{multicol}
\usepackage{caption}
\usepackage{subfigure}
\usepackage{enumitem}
\usepackage{float}
\usepackage{makecell}
\usepackage{tabu}

\usepackage[T1]{fontenc}
\usepackage[utf8]{inputenc}
\usepackage{longtable}
\usepackage{color}
\usepackage{balance} 
\usepackage{tabularx}

\definecolor{mygray}{gray}{0.6}

\usepackage{tcolorbox}
\usepackage{cleveref}
\usepackage{bbm}
\usepackage{wrapfig}

\newcommand{\ours}{{\textsc{Saracoder}}}

\definecolor{right}{RGB}{0,128,96}
\definecolor{wrong}{RGB}{192,0,32}

\definecolor{blue}{rgb}{0,0,255} 

\ifCLASSOPTIONcompsoc
  \usepackage[nocompress]{cite}
\else
  \usepackage{cite}
\fi
\ifCLASSINFOpdf
\else
\fi
\begin{document}

\title{Enhancing Knowledge Density via Hierarchical Context Pruning for Retrieval-Augmented Repository-Level Code Completion}



\author{
Xiaohan Chen, Xiang Chen, Yu Tian, Piji Li, \\ Sheng-Jun Huang, and Songcan Chen, \textit{IEEE Senior Membe \& IAPR Fellow}
\IEEEcompsocitemizethanks{
\IEEEcompsocthanksitem{Xiaohan Chen, Xiang Chen, Piji Li, Sheng-Jun Huang and Songcan Chen are with MIIT Key Laboratory of Pattern Analysis and Machine Intelligence, College of Computer Science and Technology, Nanjing University of Aeronautics and Astronautics, 29 Jiangjun Road, Nanjing, China. 
E-mail:\{cxh\_030414, xiang\_chen, pjli, huangsj, s.chen\}@nuaa.edu.cn.}
\IEEEcompsocthanksitem{Yu Tian is with the 360 Security Capability Center, Qihoo 360. E-mail: tianyu181@mails.ucas.ac.cn}
}
}

\IEEEtitleabstractindextext{%

\begin{abstract}
Retrieval-Augmented Generation (RAG) has advanced repository-level code completion by retrieving external code snippets, API documentation, or type definitions to expand the model’s contextual awareness. However,  under the finite token budget of the context window, existing retrieval-augmented methods often saturate the budget with highly similar or even redundant code, which not only wastes tokens but also suppresses critical cross-file dependencies, preventing synergistic interaction between semantically similar reference snippets and cross-file dependency definitions. Therefore, further gains are limited by low knowledge density in the retrieved context rather than by model capacity. We propose SaraCoder, a retrieval-augmented framework that enhances knowledge density via hierarchical context pruning under a fixed context budget. A Hierarchical Feature Optimization module prunes an over-complete candidate pool in four stages. The freed tokens are then allocated to an External-Aware Identifier Disambiguator, which selectively injects only the most necessary cross-file dependency definitions. SaraCoder transforms context construction into a budget-aware optimization problem, balancing reference snippets and cross-file dependencies to build a context that is both knowledge-dense and semantically complete. Experiments on CrossCodeEval and RepoEval-Updated, across multiple backbone code LLMs, show that SaraCoder consistently outperforms strong RAG baselines such as GraphCoder and Repocoder in both code-level and identifier-level metrics, without increasing token usage, and yields synergistic gains when combined with other cross-file methods.
The code is available at \url{https://anonymous.4open.science/r/code-658D}.

\end{abstract}
\begin{IEEEkeywords}
 Retrieval-augmented generation, Code completion, Context optimization, knowledge density.
\end{IEEEkeywords}}




\maketitle

\IEEEdisplaynontitleabstractindextext

\IEEEpeerreviewmaketitle

\section{Introduction}

Code Large Language Models (Code LLMs)~\cite{CodeFill,AlphaCode,SantaCoder}, built on the Transformer architecture and trained on massive code corpora, have significantly advanced the intelligence of modern software development. These models compress vast programming knowledge into hundreds of millions of parameters and have been successfully applied in numerous real-world development scenarios. However, such internal knowledge often needs to be elicited by relevant information in the context~\cite{huang2024se}. Complex project-level development tasks require integrating a wider range of context, such as related API definitions, dependent modules, and type constraints, to help the model better understand and complete the task. In this scenario, relying only on the context surrounding the completion position is insufficient. This causes the LLM to generate completions that are overly localized, incomplete, and inaccurate, which motivates the use of Retrieval-Augmented Generation (RAG) for code completion.

Retrieval-Augmented Generation (RAG)~\cite{tang2023domainadaptivecodecompletion,ijcai2022p329,zhang-etal-2023-repocoder} overcomes this limitation by introducing an external knowledge retrieval mechanism. In this framework, an efficient retriever can dynamically fetch relevant code snippets, API documentation, or type definitions from an external codebase in real time, thereby expanding the model’s contextual awareness beyond what is directly visible in the current file. The generative model then integrates these retrieval results with the current context to produce syntactically correct, semantically consistent, and standard-compliant code suggestions. As a result, RAG is gradually becoming a foundational technology for accurate and trustworthy repository-level intelligent code completion. 
However, simply being able to retrieve a longer context or more content does not automatically translate into better completion. Prior work in dense retrieval has shown that low-quality texts with excessive noise or sparse key information are unlikely to align well with relevant queries~\cite{tamber2025qaea}. Moreover, most code LLMs operate within a fixed context window, which effectively imposes a finite token budget. As codebases grow and development tasks become increasingly complex, this budget limitation becomes more pronounced. Consequently, in repository-level code completion, the system must decide under a fixed token budget {\bf what to include, what to compress, and what to discard}. In other words, the challenge shifts from “whether to retrieve” to {\bf how to construct a knowledge-dense context} within a limited window.

\begin{figure*}
  \begin{center}
    \includegraphics[width=1\linewidth]{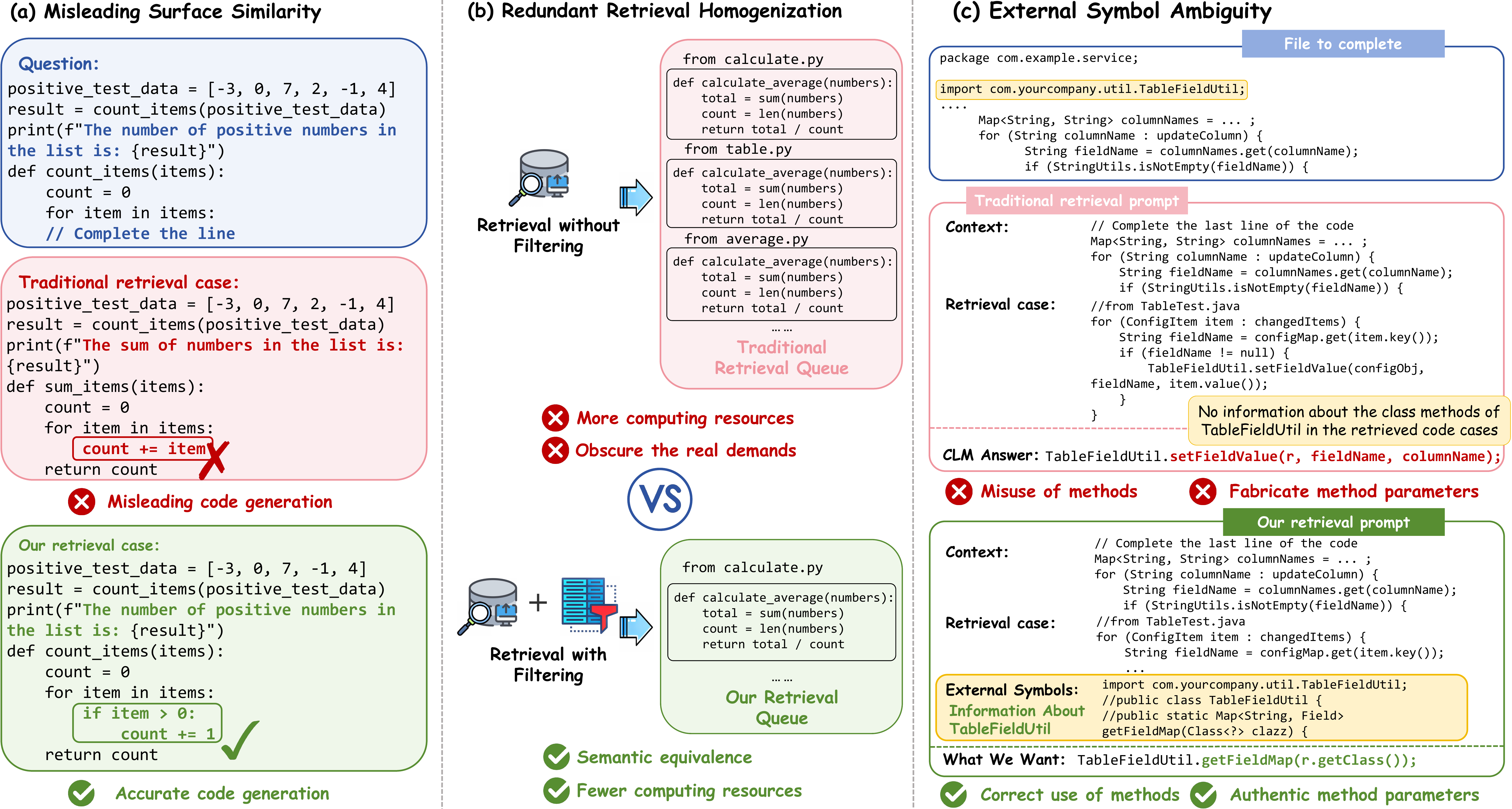}
\end{center}
\caption{The pitfalls of pure similarity retrieval vs. the efficiency of {\ours}. Pink boxes highlight how traditional retrieval leads to ``Crowding out” the context with redundant semantics or ambiguous noise. In contrast, green boxes showcase the ``Compact \& Inject” strategy of {\ours}, which distills essential knowledge and injects precise, high-value insights.}
\label{fig:background}
\end{figure*}

Current Retrieval-Augmented Generation (RAG) methods for code completion often suffer from a single-aspect knowledge focus and low knowledge density in the constructed context. Methods like GraphCoder \cite{Graphcoder} focus on retrieving similar code snippets, while others like DraCo \cite{cheng-etal-2024-dataflow} prioritize cross-file context.  This narrow focus overlooks the need to combine different types of knowledge, because developers in real-world settings require both cross-file context for project-wide consistency and similar code snippets for structural reference and efficiency.  Simply combining them makes them compete for the limited token budget, and the lack of a principled pruning strategy leads to suboptimal allocation.
In practice, this manifests in three major flaws of existing similarity-based retrieval approaches. {\bf (a) Misleading Superficial Similarity:} They often retrieve irrelevant code snippets that share surface-level similarities, which can mislead the model and result in incorrect or ineffective suggestions. {\bf (b) Redundant Retrieval Homogenization:} Relying solely on similarity ranking frequently yields redundant or duplicate snippets. This wastes valuable context window space and provides a homogeneous, narrow viewpoint, lowering the effective knowledge density of the retrieved context and hindering the model’s ability to generate innovative or superior solutions. {\bf (c) External Symbol Ambiguity:} These methods fail to capture essential dependencies like unreferenced classes or architectural rules. This ambiguity can trigger a cascade of errors, including failed type inference and mismatched signatures, compromising the correctness and reliability of the generated code. 
These flaws degrade retrieval quality and worsen the finite context budget problem (Figure~\ref{fig:background}). Current methods fill the window with low-quality snippets, crowding out critical dependencies.

To address these challenges, we introduce {\ours}, a retrieval-augmented framework that enhances knowledge density via hierarchical context pruning for repository-level code completion. It treats the fixed context window as a strict token budget and optimizes how this budget is allocated between “reference code” and “cross-file dependencies”. Instead of simply adding more similar snippets, {\ours} performs multi-stage pruning on an over-complete candidate pool of retrieved code, and then uses the saved budget to inject the most necessary dependencies.

Concretely,  {\ours}'s Hierarchical Feature Optimization module performs multi-stage hierarchical context pruning: it distills deep semantic relationships, removes exact textual duplicates, measures graph-based structural similarity, and finally applies diversity-aware reranking.  This process compresses the raw retrieval results into a compact reference set with higher knowledge density and diversity, reducing token usage. The freed budget is then used by an External-Aware Identifier Disambiguator, which resolves external symbol ambiguity by analyzing project-level dependencies and injecting cross-file definitions that are most necessary for the current completion. In this way, {\ours} turns context construction into a budget-aware optimization problem. Rather than only maximizing similarity or dependency coverage, it optimizes the joint composition of the context under a fixed token limit, so that every token in the window contributes as much useful knowledge as possible to the completion model. Our key contributions are:

(1) We propose {\ours}, a framework that enhances code completion by pruning redundant context and reallocating tokens to key cross-file dependencies, enabling informative repository-level completions within fixed budgets.

(2) {\ours} is an intelligent knowledge compressor. As merely adding similar examples shows limited gains, it compresses retrieved candidates via density-aware filtering based on semantics and structure, advancing retrieval from simple matching to intelligent, budget-aware optimization.

(3) {\ours} shows synergistic gains with existing methods. By removing redundancy and injecting critical dependencies, it complements other techniques orthogonally, minimizing interference, maximizing the preservation and utility of external cross-file knowledge and improving performance across multiple LLMs and benchmarks.


\section{Related Work}

\subsection{Reference Code Retrieval Augmentation}
This approach enhances the quality of code LLM generation by retrieving semantically reference code snippets and integrating them into prompts, mimicking the reference behaviors of programmers. CodeSearchNet~\cite{codesearch} pioneers large-scale code corpora construction, providing retrieval-based completion references; CodeRetriever~\cite{coderetriever} integrates pretrained models like CodeBERT~\cite{feng-etal-2020-codebert} to enhance complex scenario handling; ReACC~\cite{reacc} combines vector and string-based retrieval to significantly optimize long-code processing; 
GraphCoder~\cite{Graphcoder} improves code completion by using program dependencies for structured representations, allowing coarse-to-fine retrieval for Python and Java. However, like many other methods, its dependency analysis does not fully grasp deep semantic relationships in code. Additionally, most approaches rely too much on surface-level textual similarity.  This often results in redundant retrieved content, wasting resources.

\subsection{Cross-file Dependency RAG}
This method approaches code completion in complex repositories by leveraging cross-file code context (e.g., dependencies, dataflow).  Inspired by Ding et al.'s~\cite{crosscodeeval} observation that subsequent content of high-similarity snippets effectively informs completion, it injects these snippets into prompts. COCOMIC~\cite{ding-etal-2024-cocomic} dynamically fuses the context of this file with the cross-file entities retrieved by CCFinder (compressed into [SUM] vectors) through a joint attention mechanism to achieve location-aware code completion. DraCo~\cite{cheng-etal-2024-dataflow} extends this paradigm through dataflow-guided retrieval, parsing private repositories into code entities and constructing a repository-specific context graph reflecting dependencies.  DraCo retrieves precise contextual knowledge from this graph to generate well-structured prompts, overcoming cross-file information barriers and repository-specific accuracy gaps.  Current limitations include Python-exclusive implementation with type-sensitive dependencies lacking multilingual support.

\subsection{Repository-Level Evaluation}
Traditional code completion benchmarks like~\cite{HumanEval,MBPP} focus on isolated snippets, but modern software development's complexity demands more realistic evaluation.  Specialized benchmarks such as RepoEval~\cite{zhang-etal-2023-repocoder}, CrossCodeEval~\cite{crosscodeeval}, RepoBench~\cite{liu2024repobench}, and ReccEval~\cite{cheng-etal-2024-dataflow} have thus emerged, providing standardized, rigorous tests across various languages and project scales.  These benchmarks assess repository-level code completion from multiple dimensions, covering both common scenarios that rely solely on the current file's context~\cite{zhang-etal-2023-repocoder} and more complex tasks that require understanding and completing code with dependencies on symbols from other files~\cite{crosscodeeval,cheng-etal-2024-dataflow}.  They mark a shift from simple, isolated testing toward comprehensive evaluation that reflects real‑world coding environments.

\section{Problem Definition}
\label{sec:Analysis}
Given a code repository $\mathcal{R}$ and a current edit point (i.e., a target position $p$ in a target file $f_t$), the task of repository-level code completion is to generate the subsequent code sequence $C_{next}$. The context $\mathcal{C}$ on which the model relies is constructed from relevant knowledge retrieved from $\mathcal{R}$, and is strictly constrained by a predefined maximum length $L_{max}$ (in tokens), i.e., $|\mathcal{C}| \leq L_{max}$.
A well-informed context $\mathcal{C}$ should consist of two complementary components:
A set of reference code snippets $\mathcal{S}_{ref}$: A collection of code snippets retrieved from $\mathcal{R}$ that are functionally or logically similar to the current editing task. They provide the model with implementation examples and programming patterns.
A set of cross-file dependency definitions $\mathcal{D}_{ext}$: A collection of code definitions for cross-file symbols (e.g., functions, classes, variables) that are referenced (or likely to be referenced) by the current edit point and are defined within $\mathcal{R}$. They provide the model with necessary API semantics and constraints.
Thus, $\mathcal{C} = \mathcal{S}_{ref} \oplus \mathcal{D}_{ext}$, where $\oplus$ denotes concatenation.

We observe that the $\mathcal{S}_{ref}$ obtained through standard retrieval methods often suffers from high redundancy (e.g., nearly identical code snippets being retrieved repeatedly) and superficial similarity with semantic or functional ambiguity. This causes $\mathcal{S}_{ref}$ to consume a large number of tokens while providing repetitive knowledge. The consequence is that the space left for $\mathcal{D}_{ext}$, i.e., $L_{max} - |\mathcal{S}_{ref}|$, is severely compressed, forcing the system to discard many critical dependency definitions. To address this, knowledge density is introduced, which refers to the concentration of effective, non-redundant semantic knowledge within a unit of carrier (e.g., token) during knowledge transmission. We formally define the core problem as maximizing the generation quality $Q$ by optimizing the knowledge density of $\mathcal{S}_{ref}$ and the necessity of $\mathcal{D}_{ext}$:
\begin{equation}
\max_{\mathcal{S}_{ref}, \mathcal{D}_{ext}} Q(M, \mathcal{C}=\mathcal{S}_{ref} \oplus \mathcal{D}_{ext})
~\text{s.t. } |\mathcal{S}_{ref}| + |\mathcal{D}_{ext}| \leq L_{max}
\end{equation}
Consequently, the optimal reference set and external dependencies are formulated as follow. Here, $\mathcal{I}(\cdot)$ quantifies the knowledge density of a code snippet set, ensuring the selected set $\mathcal{S}$ minimizes redundancy while maximizing semantic coverage. And $\mathcal{N}(\cdot)$ measures the necessity of external dependencies.
\begin{equation}
    \mathcal{S}_{ref}^* = \arg\max_{\mathcal{S} \subseteq \mathcal{P}} \mathcal{I}(\mathcal{S}) \quad \text{s.t.} \quad |\mathcal{S}| \leq L_1
\end{equation}
\begin{equation}
        \mathcal{D}_{ext}^* = \arg\max_{\mathcal{D} \subseteq \mathcal{R}} \mathcal{N}(\mathcal{D} | \mathcal{S}_{ref}^*) \quad \text{s.t.} \quad |\mathcal{D}| + |\mathcal{S}_{ref}^*| \leq L_{max}
\end{equation}

\section{Method}
\label{sec:method}

\begin{figure*}
\begin{center}
\includegraphics[width=\linewidth]{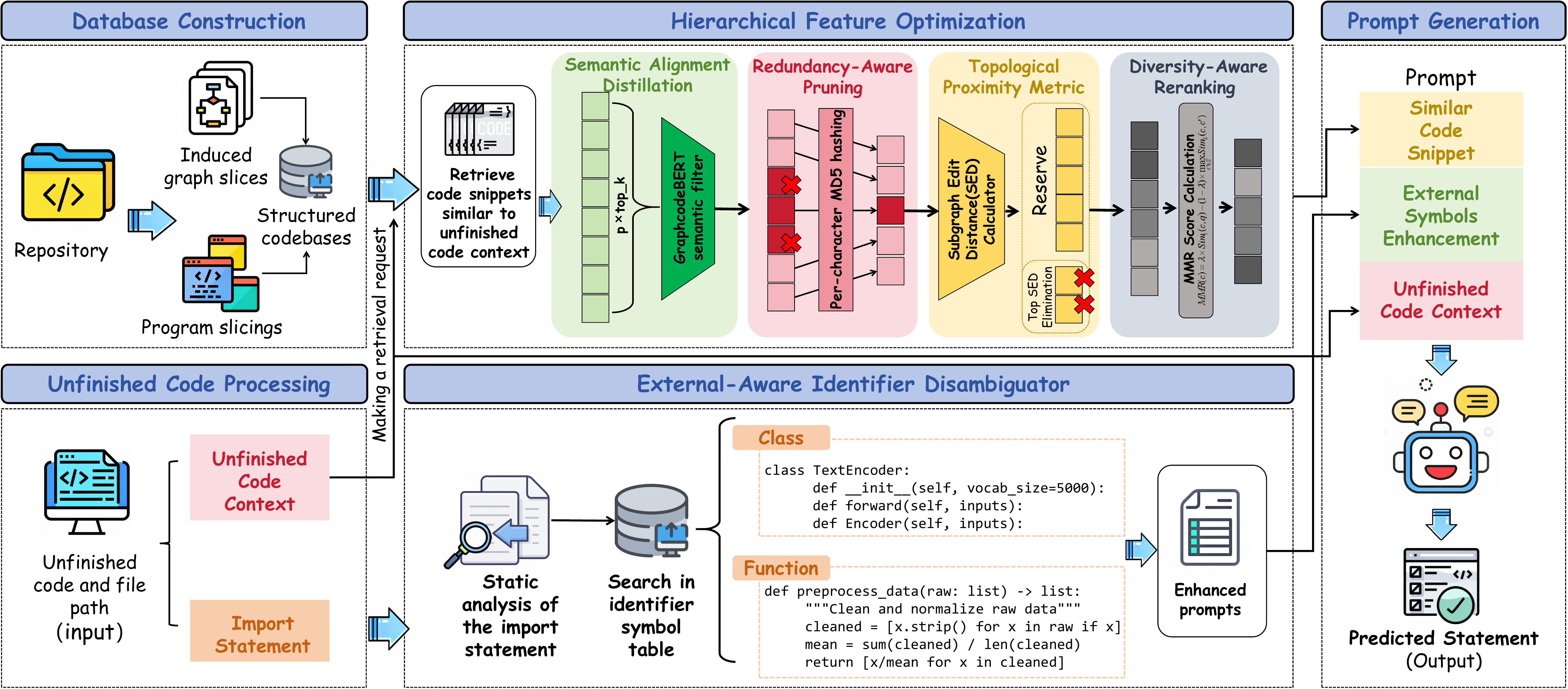}
\end{center}
\caption{An illustration of {\ours} framework.
\textbf{(1) Database Construction.} This phase constructs a key-value codebase. This involves using a slicing algorithm to create induced graph slices, which are then precisely mapped to source code snippets. \textbf{(2) Code Retrieval.} This phase takes code context as input and retrieves similar code, then refines suggestions via Hierarchical Feature Optimization. Concurrently, an External-Aware Identifier Disambiguator clarifies external symbols via dependency analysis, delivering highly accurate candidates.  \textbf{(3) Code Generation.} This phase generates prompts by integrating outputs from code retrieval with the code completion context. These prompts are then fed into an LLM to predict completion statements.
}
\label{fig:method}
\end{figure*}

The framework of {\ours} is illustrated in Figure~\ref{fig:method}, whose core objective is to intelligently allocate the budget under the hard constraint of $L_{max}$ to build a balanced and informative context $\mathcal{C}$.

\subsection{Database and Initial Candidate Construction}
We construct a searchable codebase using a multi-level context graph model that integrates control flow, data, and control dependencies~\cite{Graphcoder}. This structured representation provides stronger generalization capability than serialization-based methods. Through program slicing, task-relevant subgraphs are extracted from source code to form this database.
For each code completion query, we extract the Unfinished Code Context $C_{\text{context}}$ and the Import Statements $I$. The context $C_{\text{context}}$ is then used to retrieve an initial set of candidate code snippets, which constitutes the preliminary candidate pool $\mathcal{P}$, via text similarity search. The pool size is set as $|\mathcal{P}| = \mathit{top\_k} \times k$ where $k > 1$, ensuring a sufficiently large and diverse input for Hierarchical Feature Optimization stage. This expanded candidate set $\mathcal{P}$ serves as the direct input to HF\_OP, where it will be refined into a compact, high-quality reference set $\mathcal{S}_{\mathit{ref}}$.

\subsection{Hierarchical Feature Optimization as Context Pruning}
\label{sec:He_op}
The input of Hierarchical Feature Optimization (HF\_OP) is the preliminary candidate pool $\mathcal{P}$, and its goal is to output a refined, high-knowledge-density reference code set $\mathcal{S}_{ref}$, where the average token count of $|\mathcal{S}_{ref}|$ is much smaller than that of $|\mathcal{P}|$. This process consists of four sequentially executed sub-modules, each designed to eliminate a specific type of redundancy or noise.

\textbf{Phase 1: Semantic Alignment Distillation.}
This initial phase uses GraphcodeBERT~\cite{guo2021graphcodebert} to ensure that the candidates are highly aligned with the intent of the current query, thus filtering out irrelevant semantic noise.
First, the query code $Q$ and candidate pool $\mathcal{P}$ are tokenized into subword sequences and uniformly padded or truncated to a fixed length $L=512$. Subsequently, during the feature encoding phase, a $768$-dimensional semantic vector $v_s$ is extracted for each code unit $s \in {Q} \cup \mathcal{P}$, with vector space standardized through L2 normalization. When code repositories lack sufficient repetitive or relevant code, standard filtering methods are too strict, often leading to \textbf{zero-candidate scenarios}. This scarcity of reference material then hurts the accuracy of large language models. To fix this ``one-size-fits-all” problem, we introduce a dynamic quantile threshold mechanism. During the dynamic filtering phase, the cosine similarity set $S = \{\cos(v_Q, v_c) \mid c \in \mathcal{P}\}$ is computed between the query vector $v_Q$ and all candidate vectors $v_c$. An adaptive threshold $\tau = \text{quantile}(S, 0.75)$ is set at the 75th percentile, outputting filtered results $C_{SAD} = \{c \mid \cos(v_Q, v_c) \geq \tau\}$. To reduce redundant computation overhead and improve efficiency, a caching mechanism stores encoding results for high-frequency code.

\textbf{Phase 2: Redundancy-Aware Pruning.}
This phase implements lightweight hash-based deduplication via exact text matching. Using the MD5 algorithm~\cite{10.17487/RFC1321}, it generates 128-bit hash fingerprints (\emph{single computation} $\approx 0.02$ms, \emph{memory footprint} $32$ bytes/hash) to eliminate verbatim duplicates from candidate set $C_{SAD}$ with minimal computational cost, significantly reducing downstream overhead. The module maintains a global hash set $H_{\text{seen}}$ to dynamically track processed sample fingerprints: for each candidate $c \in C_{SAD}$, if its MD5 hash $h_c \notin H_{\text{seen}}$, $c$ is added to deduplicated result set $C_{RAP}$ and $H_{\text{seen}}$ is updated. This achieves processing with $O(N)$ time complexity. After semantic alignment distillation processing, the number of code snippets requiring MD5 hashing is limited and their structure is fixed by syntactic and semantic constraints.  The MD5 collision resistance (theoretical probability $\approx 1.47 \times 10^{-18}$) is sufficient for strict sensitivity. Additionally, MD5's superior speed and lower memory footprint provide optimal cost-performance. Overall, this lightweight hash-based pruning stage aggressively removes verbatim duplicates at small computational cost, freeing token budget from exact redundancy and further concentrating knowledge density in the surviving candidates.

\textbf{Phase 3: Topological Proximity Metric.}
At this phase, the decaying subgraph edit distance (D-SED) is introduced to measure the graph similarity between the query graph $G_q$ and the candidate graph $G_c$ (\cite{ranjan2022greed,graphED}). We calculate D-SED for code snippets to quantify their structural similarity and retain those with the closest match. Editing operations \(O\) are the set of operations to transform $G_c$ to $G_q$, include adding, deleting, and modifying nodes or edges.  Each operation $op \in O$ has a cost \(c(op)\) and a hop count \(l(op)\) from its ``core node”. For simplicity, we choose the node with the largest ID as core node. Operations closer to the core exert greater structural influence. \(\gamma \in (0,1)\) is an attenuation factor that reduces the cost weight for operations farther from the core node.
\begin{equation}
    D-\operatorname{SED}\left(G_{q}, G_{c}\right)=\sum_{o p=\mathrm{O}} \gamma^{l(o p)} \cdot c(o p)
\end{equation}
By prioritizing structural proximity, this phase prunes structurally irrelevant or overly distant snippets. After computing D-SED scores for each candidate $c \in C_{RAP}$, we compute a composite score $s$ as a weighted sum of text similarity (calculated in initial candidate generation) and structural similarity (D-SED scores).  We generate $Q_{TPM} = [(\text{c}, \text{s}), \ldots]$, ordered in descending score $s$. 

\textbf{Phase 4: Diversity-Aware Reranking.}
This module implements a variation-aware ranking model based on the Maximal Marginal Relevance (MMR)~\cite{MMR} algorithm, designed not only to enhance diversity but also to maximize the density of semantic knowledge per token. By employing adversarial similarity measurement and dynamic weight adjustment, the model counters the homogeneity typical of conventional ranking approaches. Here, $S$ consists of items $(c,s) \in Q_{TPM}$ not yet chosen for $\mathcal{S}_{ref}$. $\text{Sim}_1$ captures the relevance ($s_i = \pi_2 \circ \iota{c_i}(S)$) of candidate $c_i$ to query $q$, while $\text{Sim}_2$ denotes the maximum cosine similarity between $c_i$ and any item $c_j$ in the selected set $\mathcal{S}_{ref}$. The trade-off parameter $\lambda$ balances relevance ($\text{Sim}_1$) against diversity ($\text{Sim}_2$), ultimately optimizing the knowledge density of the ranked output.
\begin{equation}
\text{MMR} = \arg\max_{c_i \in S} \Bigl[ \lambda \cdot \text{Sim}_1(c_i, q) - (1 - \lambda) \cdot \max_{c_j \in \mathcal{S}_{ref}} \text{Sim}_2(c_i, c_j) \Bigr]
\end{equation}
By iteratively penalizing candidates similar to already selected, this MMR-based stage prunes away homogeneous snippets that would consume additional tokens without adding new signals.  Combined with the previous phases, HF\_OP thus implements a hierarchical context pruning pipeline that transforms an oversized, redundant candidate pool $\mathcal{P}$ into a small, knowledge-dense reference set $\mathcal{S}_{\mathit{ref}}$ suitable for use under a strict context budget.

\begin{figure}
\begin{center}
\includegraphics[width=1\linewidth]{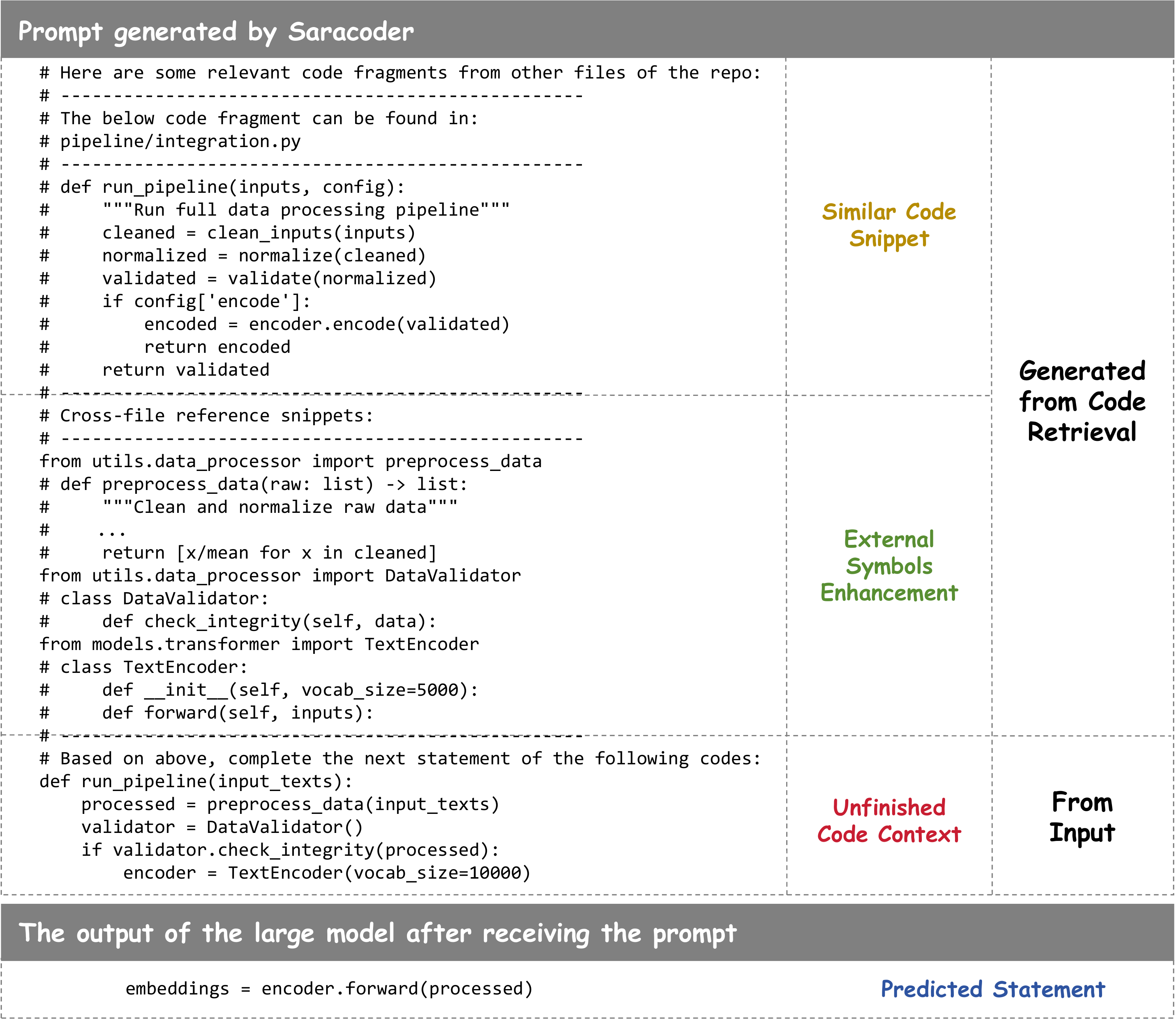}
\end{center}
\caption{Prompt template. ($\mathcal{S}_{ref}$ corresponds to similar code snippet, $\mathcal{D}_{ext}$ corresponds to external symbols enhancement, and $C_{\text{context}}$ corresponds to unfinished code context.)}
\label{fig:prompt_fram}
\end{figure}

\subsection{External-Aware Identifier Disambiguator}
By utilizing the tokens saved in Section~\ref{sec:He_op}, this module is able to inject cross-file identifier knowledge to enhance the model's knowledge. Firstly, file-level entity modeling parses code per file $F$, extracting method entities $E_{\text{method}}$ (functions/class methods with identifier, alias, line range $[l_{\text{start}}, l_{\text{end}}]$, parameter signature, scope) and class entities $E_{\text{class}}$ (class definitions with identifier, alias, line range $[l_{\text{start}}, l_{\text{end}}]$, member mappings) that built in the $F$. After that, it generates a structured identifier symbol table $ST_{\text{lib}} = \{ \text{identifier} \mapsto \text{ syntax features} \} $, where $\mathit{identifier}$ corresponds to either: (1) the unique identifier of a method entity $\forall e \in E_{\text{method}}$, or (2) the unique identifier of a class entity $\forall c \in E_{\text{class}}$, with the mapped $\mathit{syntax\ features}$ containing all associated attributes for that entity.
Subsequently, the dependency resolution mechanism processes all import statements ($I$) within the unfinshed file. For Intra-project Cross-module Reference, this phase retrieves complete entities ($E_{lib}$) from the pre-built entity library ($ST_{\text{lib}}$) by determining their corresponding file paths ($p$). These paths are constructed through decomposition of module components derived from either dotted names (e.g., \texttt{my.module.MyClass}) or relative imports (e.g., \texttt{from .sub\_module import MyClass}), which are subsequently joined using directory separators (\texttt{/}) and appended with the \texttt{.py} file extension. For standard and third-party libraries, the system constructs a lightweight reference table $T_{\text{ext}} = \{ \text{canonical name} \mapsto \text{alias} \}$ to efficiently manage cross-file dependencies without full entity resolution. The final output $\mathcal{D}_{ext} = I \oplus E_{lib} \oplus T_{\text{ext}}$.

\subsection{Prompt Generation}
\label{sec:prompt}
Following code retrieval and external link resolution, {\ours} employs an external LLM to generate subsequent statements. The final prompt $\mathcal{C}$ is constructed by concatenating three components:
\textbf{cross-file dependency definitions} $\mathcal{D}_{ext}$, \textbf{reference code snippets} $\mathcal{S}_{ref}$ and \textbf{unfinished code context} $C_{\text{context}}$. This architecture follows $\mathcal{C} = \mathcal{S}_{ref} \oplus \mathcal{D}_{ext} \oplus  C_{\text{context}}$. (As shown in Figure~\ref{fig:prompt_fram}).

\section{Experiments}

\label{sec:experiments}

\subsection{Experimental Setup}
\subsubsection{Datasets}

We primarily utilize two datasets here: \textbf{(1) CrossCodeEval~\cite{crosscodeeval}:} This benchmark evaluates code completion in complex scenarios like type inference and dependency analysis. It is ideal for assessing performance that requires a deep understanding of code across multiple files. \textbf{(2) RepoEval-Updated~\cite{Graphcoder}:} The dataset extends the original RepoEval~\cite{zhang-etal-2023-repocoder} by adding Java tasks. This new version includes repositories of varying scales, offering a better way to assess basic syntax, common API usage, and local context understanding. Table~\ref{tab:dataset} shows details.

\begin{table}
    \centering
    \caption{CrossCodeEval vs. RepoEval-Updated.}
    \resizebox{\linewidth}{!}{
    \begin{tabular}{ccccc}
    \hline
    \toprule
    & \multicolumn{2}{c}{\textbf{CrossCodeEval}} 
    & \multicolumn{2}{c}{\textbf{RepoEval-Updated}} 
    \\
    \cmidrule(lr){2-3}\cmidrule(lr){4-5}
    &\textbf{Python} & \textbf{Java}
    & \textbf{Python}&\textbf{Java}
    \\    
    \midrule[0.5pt]
    Total Repositories & 471 & 239 & 10 & 8
    \\
    Total Files & 14348 & 5868 &  3258 & 8260
    \\
    Total Task cases & 2665 & 2139 & 2000 & 1600
    \\
    \bottomrule
    \end{tabular}
    }
    \label{tab:dataset}
\end{table}

\subsubsection{Evaluation Indicators Setting}
In this study, the following several evaluation indicators are used to assess the effect of code completion~\cite{computeEMES,crosscodeeval}.
\textbf{(1) Code Exact Match (EM):} Proportion of generated code exactly matching the ground truth. 
\textbf{(2) Identifier Exact Match (ID\_EM):} The percentage of identifiers (variables, functions, etc.) perfectly matching the ground code. 
\textbf{(3) Identifier F1 Score (ID\_F1):} A more nuanced evaluation of identifier matching by combining precision and recall. It offers a more comprehensive assessment of identifier completion quality.
\textbf{(4) Edit Similarity (ES):} Similarity metric between generated and ground-truth code based on edit distance. It tolerates slight variations.

\subsubsection{Model Selection and Baseline Setting}
In this experiment, we select  Codegen2-7b,  Codegen25-7b~\cite{codegen,codegen2},  CodeLlama-7b-Instruct~\cite{codellama} and  deepseek-coder-6.7b-instruct~\cite{deepseekcoder} for code completion task inference. We employ the following five methods as controls to evaluate the effectiveness of retrieval-augmented generation (RAG) in code completion: 
\textbf{(1) No RAG:} Relies solely on the LLM’s pre-trained knowledge for autoregressive generation given the prefix code, without any retrieval from the codebase. Serves as a zero‑shot baseline.
\textbf{(2) Shifted RAG:} Employs a sliding‑window offset mechanism, which dynamically adjusts window positions during retrieval, prioritizing code segments likely to contain target call chains.  It enhances temporal relevance through temporal probability prediction, showing strength in scenarios like API invocation sequences and control flow continuation.
\textbf{(3) Vanilla Rag:} Given the context, retrieve a set of similar code snippets from the repository through a fixed-size sliding window and call the LLM to obtain the predicted next statement.
\textbf{(4) Repocoder~\cite{zhang-etal-2023-repocoder}\footnote{
We attempt to run the code publicly released by Zhang et al\cite{zhang-etal-2023-repocoder}, but fail to execute them with the provided instructions. Our specific implementation here is based on the work of CrossCodeEval~\cite{crosscodeeval}. }:}  An iterative retrieval‑augmented framework for repository‑level completion. It integrates similarity‑based retrieval with pretrained code LLMs to enable precise cross‑file completion.
\textbf{(5) Graphcoder~\cite{Graphcoder}:} A structured retrieval‑generation framework that uses Code Context Graphs to model code dependencies, replacing conventional sequence‑based context representations.

\subsection{Experimental Details}
\subsubsection{Context Graph Construction} 
Code parsing transforms source code into an intermediate representation that is easier to analyze and process, and is a fundamental step to understand the semantics and structure of a program. Abstract Syntax Tree is a widely used and effective intermediate representation, mapping source code to a tree topology structure, and accurately representing the syntax features and context relationships of code elements. By traversing and manipulating the abstract syntax tree, the relationship graph in the code can be constructed efficiently. Tree-sitter is a CFG-based parser generator that can support a variety of programming languages including Python, Java, C++. The core advantage of Tree-sitter is its efficient parsing performance and wide support for multiple languages. In code, the following relationships play a key role in semantic analysis, refactoring, debugging, and maintenance. We use tree-sitter to model these. See Table~\ref{tab:semantic-relationships}.

\begin{table*}[ht]
\centering
\caption{Key Associations for Analysis, Refactoring, and Debugging Modeled in Context Graph Construction}
\setlength{\tabcolsep}{5pt}
\renewcommand{\arraystretch}{1.15}
\renewcommand{\tabularxcolumn}[1]{m{#1}}
\begin{tabularx}{\textwidth}{@{}l >{\raggedright\arraybackslash}X >{\raggedright\arraybackslash}X@{}}
\toprule
\textbf{Relationship} & \textbf{Definition and Type-Sensitive Characteristics} & \textbf{Syntax Examples (Python / Java)} \\
\midrule
Assignment & A variable acquires its type identity through assignment, which enables type inference and propagation. & 
\makecell[l]{\textbf{Python:} \texttt{count: int = 10}; \\ \textbf{Java:} \texttt{String s = new String()}.} \\
\addlinespace

Contextual Binding & Temporary type bindings are created within specific syntactic structures, and their lifecycle is managed depending on the context. & 
\makecell[l]{\textbf{Python:} \texttt{with open(file) as f:}; \\ \textbf{Java:} \texttt{try (BufferedReader br = ...)}.} \\
\addlinespace

Reference & Existing variables or properties are accessed, which requires late-bound type resolution. & 
\makecell[l]{\textbf{Python:} \texttt{obj.calculate()}; \\ \textbf{Java:} \texttt{this.value}.} \\
\addlinespace

Type Declaration & Variable or return types are explicitly annotated, which provides the basis for static type checking. & 
\makecell[l]{\textbf{Python:} \texttt{def func() -> list[str]:}; \\ \textbf{Java:} \texttt{List<Integer> list = new ArrayList<>()}.} \\
\addlinespace

Parameter Constraint & Function parameters are constrained by types, which ensures input type validation. & 
\makecell[l]{\textbf{Python:} \texttt{def sort(items: Sequence[T])}; \\ \textbf{Java:} \texttt{void sort(List<? extends Comparable> l)}.} \\
\addlinespace

Return Constraint & Function return values are constrained by types, which guarantees output type consistency. & 
\makecell[l]{\textbf{Python:} \texttt{@return\_type(float)}; \\ \textbf{Java:} \texttt{public int getValue() \{ ... \}}.} \\
\addlinespace

Inheritance & Subclasses automatically acquire members from their parent class, forming a type hierarchy. & 
\makecell[l]{\textbf{Python:} \texttt{class Child(Parent):}; \\ \textbf{Java:} \texttt{class Child extends Parent \{ ... \}}.} \\
\addlinespace

Implementation & A class fulfills an interface contract, which serves as the foundation for polymorphic behavior. & 
\makecell[l]{\textbf{Python:} \texttt{class MyList(ABC):}; \\ \textbf{Java:} \texttt{class ArrayList implements List \{ ... \}}.} \\
\addlinespace

Override & A subclass overrides a parent class method, which enables dynamic method dispatch. & 
\makecell[l]{\textbf{Python:} \texttt{def method(self): ...}; \\ \textbf{Java:} \texttt{@Override void method() \{ ... \}}.} \\
\addlinespace

Import Dependency & Cross-module imports create dependencies, which control type visibility. & 
\makecell[l]{\textbf{Python:} \texttt{import pandas as pd}; \\ \textbf{Java:} \texttt{import java.util.List}.} \\
\addlinespace

Invocation & Methods or functions invoke each other, which requires type compatibility verification. & 
\makecell[l]{\textbf{Python:} \texttt{math.sqrt(x)}; \\ \textbf{Java:} \texttt{Collections.sort(list)}.} \\
\bottomrule
\end{tabularx}
\label{tab:semantic-relationships}
\end{table*}

\subsubsection{Parameter Setting}
In this study, we mainly adopt the Greedy Decoding strategy for text generation. Its core configuration is to disable sampling (\texttt{do\_sample=False}) and set the temperature value (\texttt{temperature}) to 0. This combination ensures that the model necessarily selects the token with the highest logical probability at each step, thereby eliminating randomness in the generation process and facilitating a strict and reproducible evaluation of the model's performance. To simplify and observe the performance of various methods within the limited length of the context window, the maximum length limit for text is 2048 new tokens (\texttt{max\_num\_tokens}).

\subsubsection{Resource Control and Allocation}
In practice,  \texttt{max\_num\_tokens} controls the total length of the context window. The system sets a maximum length limit for the ``unfinished code context,” which is no more than half of the total window. The remaining context window is dynamically used to accommodate other knowledge, including retrieved similar code snippets. Since the output of the External Symbols Enhancement module is highly compressed, we primarily use the \textit{top\_k} parameter to control the number of similar code snippets that can be included in the prompt. This effectively constrains the use of this portion of the resources. The value of \textit{top\_k} directly determines the number of similar examples that can be introduced; a smaller value allocates less context space to similar code snippets. Through this mechanism, we can achieve flexible and precise control over the different components of the knowledge within a limited total resource budget.


\begin{table*}[ht]
\centering
\caption{Performance comparison on RepoEval-Updated and CrossCodeEval datasets. Numbers are shown in percentage (\%). The top results are \textbf{bolded}, and the second best are \underline{underlined}.}
\label{tab:combined_results}
\resizebox{\linewidth}{!}{%
\begin{tabular}{clcccccccccccccccccc}
\hline
\toprule
\multicolumn{1}{c}{\multirow{4}{*}{\textbf{Dataset}}} & 
\multicolumn{1}{c}{\multirow{4}{*}{\textbf{Language}}} & 
\multicolumn{1}{c}{\multirow{4}{*}{\textbf{Methods}}} &
\multicolumn{16}{c}{\textbf{Models and Metrics}} \\
\cmidrule(l){4-19}
& & &
\multicolumn{4}{c}{\textbf{Codegen2-7b}} & \multicolumn{4}{c}{\textbf{Codegen25-7b}} & 
\multicolumn{4}{c}{\textbf{deepseek-coder-6.7b-instruct}} & \multicolumn{4}{c}{\textbf{CodeLlama-7b-Instruct}} \\
\cmidrule(lr){4-7} \cmidrule(lr){8-11} \cmidrule(lr){12-15} \cmidrule(lr){16-19}
& & &
\multicolumn{2}{c}{\textbf{Code Match}} & \multicolumn{2}{c}{\textbf{Identifier Match}} & 
\multicolumn{2}{c}{\textbf{Code Match}} & \multicolumn{2}{c}{\textbf{Identifier Match}} &
\multicolumn{2}{c}{\textbf{Code Match}} & \multicolumn{2}{c}{\textbf{Identifier Match}} & 
\multicolumn{2}{c}{\textbf{Code Match}} & \multicolumn{2}{c}{\textbf{Identifier Match}} \\
\cmidrule(lr){4-5} \cmidrule(lr){6-7} \cmidrule(lr){8-9} \cmidrule(lr){10-11}
\cmidrule(lr){12-13} \cmidrule(lr){14-15} \cmidrule(lr){16-17} \cmidrule(lr){18-19}
& & & EM & ES & EM & F1 & EM & ES & EM & F1 & EM & ES & EM & F1 & EM & ES & EM & F1 \\
\midrule[0.5pt]

\multirow{12}{*}{RepoEval-Updated} & 
\multirow{6}{*}{Python} & No Rag 
& 17.40 & 32.54 & 23.75 & 30.21 & 19.55 & 34.48 & 25.75 & 32.16 & 11.50 & 30.33 & 15.30 & 22.39 & 17.35 & 33.05 & 23.55 & 30.53 \\
& & Shifted Rag
& 32.70 & 59.22 & 40.10 & 55.66 & 36.45 & 61.96 & 43.20 & 58.31 & 20.90 & 42.95 & 26.50 & 38.88 & 33.90 & 60.28 & 41.50 & 56.39 \\
& & Vanilla Rag
& 38.70 & 63.58 & 46.45 & 60.43 & 42.25 & 66.26 & 48.75 & 62.79 & 22.20 & 41.48 & 27.85 & 37.58 & 40.30 & 65.03 & 47.55 & 61.06 \\
& & Repocoder
& 37.60 & 61.98 & 45.10 & 58.47 & 40.55 & 64.48 & 46.85 & 60.71 & 21.35 & 40.18 & 26.65 & 35.93 & 39.60 & 63.71 & 47.05 & 59.78 \\
& & Graphcoder
& \underline{42.40} & \underline{65.73} & \underline{49.45} & \underline{62.07} & \textbf{44.65} & \underline{67.59} & \underline{51.00} & \underline{63.82} & \textbf{28.50} & \underline{44.63} & \underline{33.35} & \underline{42.63} & \underline{43.90} & \underline{67.26} & \underline{51.15} & \underline{63.51} \\
& & {\ours}
& \textbf{42.60} & \textbf{65.92} & \textbf{50.15} & \textbf{62.61} & \underline{44.50} & \textbf{67.79} & \textbf{51.10} & \textbf{63.84} & \underline{28.25} & \textbf{46.91} & \textbf{33.45} & \textbf{42.95} & \textbf{45.00} & \textbf{68.27} & \textbf{52.00} & \textbf{63.97} \\
\cmidrule(lr){2-19}
& 
\multirow{6}{*}{Java} & No Rag 
& 6.55 & 16.84 & 9.15 & 8.84 & 5.35 & 16.21 & 9.05 & 8.65 & 6.40 & 20.61 & 7.75 & 8.73 & 6.85 & 17.11 & 9.45 & 9.06 \\
& & Shifted Rag
& 30.87 & 62.52 & 43.94 & 61.01 & 26.63 & 58.46 & 37.75 & 56.57 & 28.00 & 55.12 & 36.81 & 53.39 & 35.38 & 64.63 & 45.56 & 62.98 \\
& & Vanilla Rag
& 33.50 & 63.82 & 45.44 & 62.08 & 32.00 & 61.52 & 41.56 & 59.48 & 21.13 & 46.51 & 31.19 & 45.15 & 38.56 & 66.46 & 48.06 & 64.90 \\
& & Repocoder
& 30.13 & 60.01 & 42.31 & 57.10 & 28.75 & 57.73 & 37.88 & 54.46 & 22.19 & 46.93 & 32.06 & 44.26 & 35.38 & 62.33 & 44.44 & 59.63 \\
& & Graphcoder
& \underline{37.75} & \underline{66.19} & \underline{50.68} & \underline{64.77} & \underline{36.63} & \underline{64.74} & \underline{46.13} & \underline{62.62} & \underline{28.81} & \underline{55.75} & \underline{40.06} & \underline{53.61} & \textbf{43.00} & \textbf{69.68} & \textbf{52.69} & \textbf{67.85} \\
& & {\ours}
& \textbf{37.93} & \textbf{67.06} & \textbf{50.93} & \textbf{65.48} & \textbf{36.75} & \textbf{65.54} & \textbf{46.44} & \textbf{62.61} & \textbf{29.75} & \textbf{56.38} & \textbf{40.75} & \textbf{54.18} & \underline{42.88} & \underline{69.37} & \underline{52.00} & \underline{67.37} \\

\midrule[0.75pt]

\multirow{12}{*}{CrossCodeEval} & 
\multirow{6}{*}{Python} & No Rag 
& 0.00 & 13.38 & 0.00 & 2.24 & 0.00 & 13.26 & 0.00 & 2.10 & 0.00 & 4.51 & 0.00 & 0.57 & 0.00 & 13.27 & 0.00 & 2.22 \\
& & Shifted Rag
& 4.84 & 46.67 & 11.48 & 42.72 & 7.40 & 48.88 & 14.09 & 44.62 & 8.19 & 50.18 & 14.77 & 46.69 & 6.91 & 49.12 & 13.60 & 45.12 \\
& & Vanilla Rag
& 9.48 & 50.97 & 17.15 & 47.81 & 12.39 & 53.92 & 23.61 & 53.14 & 13.00 & 54.00 & 26.63 & 55.77 & 11.45 & 52.93 & 19.23 & 50.49 \\
& & Repocoder
& \underline{12.47} & \underline{54.08} & \underline{21.57} & \underline{51.89} & \underline{16.62} & \underline{56.80} & \underline{25.73} & \textbf{57.85} & \underline{17.11} & \underline{58.11} & \underline{26.71} & \underline{56.46} & \underline{15.14} & \underline{56.28} & \underline{24.56} & \underline{54.22} \\
& & Graphcoder
& 10.88 & 52.36 & 19.68 & 49.73 & 14.54 & 55.29 & 23.38 & 52.61 & 15.53 & 57.05 & 24.29 & 55.01 & 13.30 & 55.41 & 22.63 & 52.91 \\
& & {\ours}
& \textbf{15.04} & \textbf{56.03} & \textbf{24.44} & \textbf{54.68} & \textbf{18.36} & \textbf{58.30} & \textbf{27.28} & \underline{56.22} & \textbf{19.72} & \textbf{59.93} & \textbf{28.52} & \textbf{58.26} & \textbf{17.91} & \textbf{58.37} & \textbf{27.77} & \textbf{56.82} \\
\cmidrule(lr){2-19}
& 
\multirow{6}{*}{Java} & No Rag 
& 1.03 & 21.79 & 0.64 & 16.86 & 1.50 & 21.77 & 24.78 & 24.78 & 6.40 & \underline{35.79} & 10.42 & 32.90 & 0.93 & 20.83 & 1.96 & 16.17 \\
& & Shifted Rag
& 6.08 & 46.09 & 12.11 & 43.76 & 5.89 & 38.00 & 10.23 & 36.44 & 5.84 & \textbf{36.19} & 11.64 & \textbf{35.23} & 6.73 & 43.75 & 12.71 & 41.68 \\
& & Vanilla Rag
& 9.30 & \textbf{47.42} & 15.71 & \textbf{45.69} & 10.38 & \underline{40.76} & 15.29 & \underline{39.91} & 8.88 & 33.59 & 15.01 & 33.51 & 10.93 & \underline{45.01} & 17.81 & \underline{44.08} \\
& & Repocoder
& \underline{10.71} & 41.83 & \underline{16.18} & 41.51 & \textbf{12.16} & \textbf{42.38} & \textbf{17.63} & \textbf{41.69} & \underline{9.58} & 34.25 & \underline{15.76} & 34.13 & \textbf{13.23} & \textbf{46.01} & \textbf{19.87} & \textbf{45.14} \\
& & Graphcoder
& 8.13 & 45.18 & 14.35 & 43.32 & 8.42 & 36.77 & 12.85 & 35.84 & 7.39 & 32.34 & 12.76 & 32.05 & 8.51 & 40.57 & 14.96 & 39.57 \\
& & {\ours}
& \textbf{11.73} & \underline{46.69} & \textbf{18.37} & \underline{45.47} & \underline{11.40} & 39.33 & \underline{16.22} & 38.97 & \textbf{11.92} & 34.55 & \textbf{17.72} & \underline{34.99} & \underline{12.95} & 42.71 & \underline{19.54} & 42.33 \\

\bottomrule
\end{tabular}
}
\end{table*}

\begin{figure*}
\begin{center}
    \includegraphics[width=\linewidth]{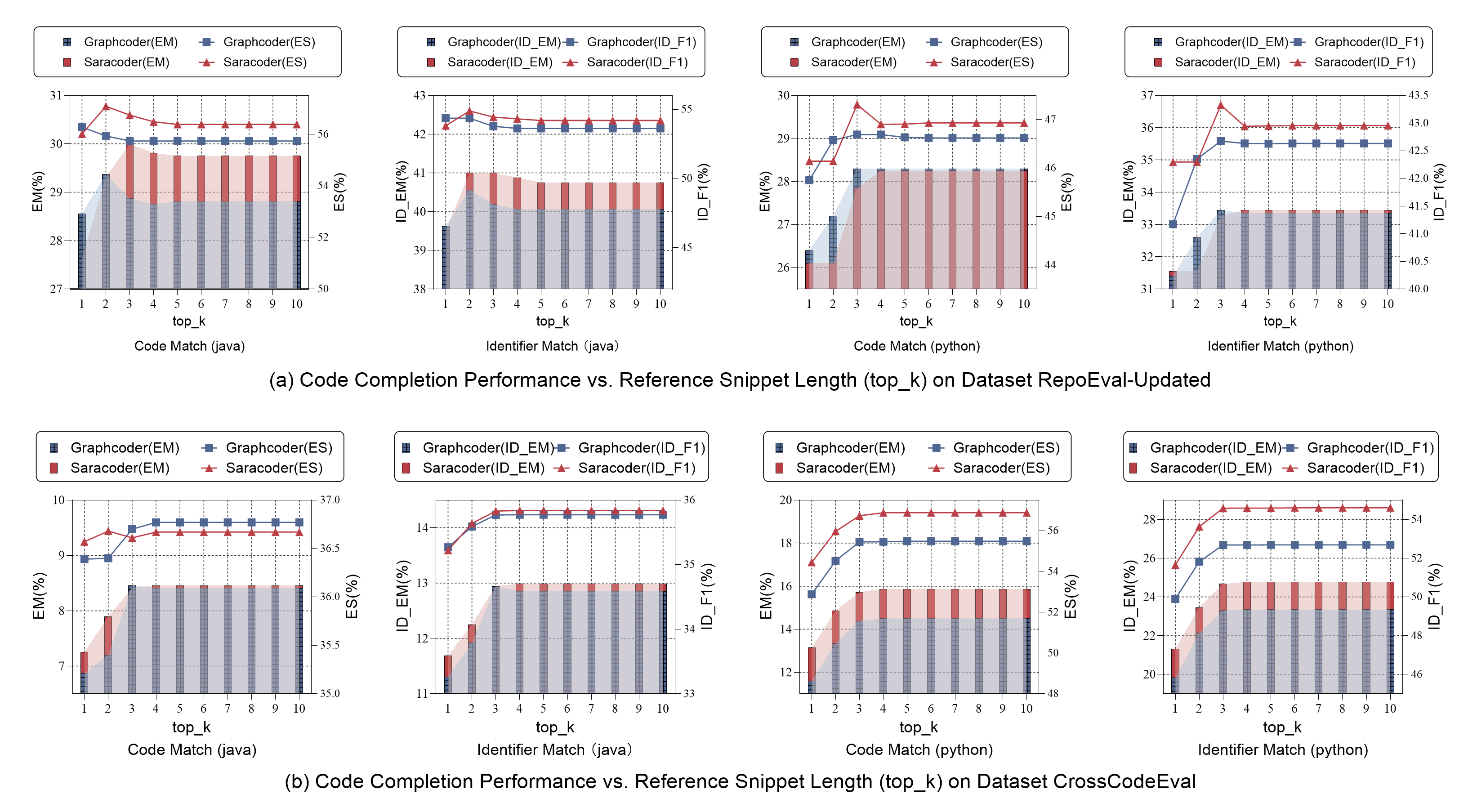}
\caption{Impact of top\_k on RepoEval-Updated and CrossCodeEval. }
\label{fig:token_cross}
\end{center}
\end{figure*}

\definecolor{softred}{RGB}{255,0,0}
\definecolor{softgreen}{RGB}{0, 128, 0}
\definecolor{softgray}{RGB}{150, 150, 150}

\begin{table*}[h]
\centering
\caption{Performance benefits of {\ours} when integrated orthogonally with other cross-file approaches ($\%$). The CrossCodeEval dataset is used in this part.
}
\resizebox{\linewidth}{!}{
\begin{tabular}{clllllllllllll}
\hline
\toprule
\multicolumn{1}{c}{\multirow{3}{*}{\textbf{Language}}}
& 
\multicolumn{1}{c}{\multirow{3}{*}{\textbf{Methods}}}
&
\multicolumn{4}{c}{\textbf{Codegen2-7b}} & \multicolumn{4}{c}{\textbf{Codegen25-7b}} 
& \multicolumn{4}{c}{\textbf{CodeLlama}}
\\
\cmidrule(lr){3-6} \cmidrule(lr){7-10} \cmidrule(lr){11-14}
& &\multicolumn{2}{c}{\textbf{Code Match}} & \multicolumn{2}{c}{\textbf{Identifier Match}} 
& \multicolumn{2}{c}{\textbf{Code Match}}&\multicolumn{2}{c}{\textbf{Identifier Match}} 
&\multicolumn{2}{c}{\textbf{Code Match}} & \multicolumn{2}{c}{\textbf{Identifier Match}} 
\\
\cmidrule(lr){3-4} \cmidrule(lr){5-6} \cmidrule(lr){7-8} \cmidrule(lr){9-10}
\cmidrule(lr){11-12} \cmidrule(lr){13-14}
& & \multicolumn{1}{l}{EM} & \multicolumn{1}{l}{ES} & \multicolumn{1}{l}{EM} & \multicolumn{1}{l}{F1} & \multicolumn{1}{l}{EM} & \multicolumn{1}{l}{ES} & \multicolumn{1}{l}{EM} & \multicolumn{1}{l}{F1} & \multicolumn{1}{l}{EM} & \multicolumn{1}{l}{ES} & \multicolumn{1}{l}{EM} & \multicolumn{1}{l}{F1}
\\
\midrule[0.5pt]
\multicolumn{1}{c}{\multirow{8}{*}{Python}}
&  No Rag & 
5.44  & 57.85 & 11.71 & 42.22 &
7.77  & 60.52 & 14.45  & 45.40  &
9.49  & 61.97 & 16.44  & 47.36 
\\
& Shift Rag &
4.87 & 58.36 & 11.64  & 42.91  &
7.44  & 60.20 & 14.17  & 44.78 & 
6.95 & 60.35 & 13.75 & 45.36 
\\
&  Vanilla Rag & 
9.52  & 61.87 & 17.42 & 48.01 & 
12.43 & 63.81 & 20.74 & 51.00  &
11.48 & 63.66& 19.42& 50.72
\\
&  {\ours} & 
12.16  & 54.16 & 18.37  & 45.47  &
11.50  & 44.90 & 16.22  & 38.95  &
13.32  & 48.04 & 19.54  & 42.34  
\\
\cmidrule(lr){2-14}
&  Repocoder & 
12.50  & 64.48 & 21.87 & 52.09 &
16.66 & 66.67 & 25.99 & 55.06  &
15.19 & 66.24& 24.74& 54.39 
 
\\
&  Repocoder + {\ours} & 
15.94 \small{ \textcolor{softgreen}{+3.44}} & 66.43 \small{\textcolor{softgreen}{+1.95}} & 25.73 \small{ \textcolor{softgreen}{+3.86}} & 54.80 \small{\textcolor{softgreen}{+2.71} }&

19.49 \small{\textcolor{softgreen}{+2.83}} & 68.53 \small{ \textcolor{softgreen}{+1.86}}& 28.90 \small{ \textcolor{softgreen}{+2.91}} & 57.49 \small{ \textcolor{softgreen}{+2.43}}&

19.19 \small{\textcolor{softgreen}{+4.00}} & 68.45 \small{ \textcolor{softgreen}{+2.21}}& 29.05 \small{ \textcolor{softgreen}{+4.31}} & 57.47 \small{ \textcolor{softgreen}{+3.08} }
\\

\cmidrule(lr){2-14}
&  Draco & 
20.06  & 66.33 & 29.13  & 56.53  &
22.93  & 68.70 & 32.45 & 59.34 &
23.50  & 68.56 & 32.57 & 59.49
\\
&  Draco + {\ours} & 
24.06 \small{ \textcolor{softgreen}{+4.00} }& 69.40 \small{ \textcolor{softgreen}{+3.07}}& 34.00 \small{ \textcolor{softgreen}{+4.87}} & 61.12 \small{\textcolor{softgreen}{+4.59}} &

27.05 \small{\textcolor{softgreen}{+4.12}} & 71.86 \small{ \textcolor{softgreen}{+3.16}}& 36.80 \small{ \textcolor{softgreen}{+4.35}} & 63.48 \small{ \textcolor{softgreen}{+4.14} }&

27.20 \small{ \textcolor{softgreen}{+3.7} }& 72.01 \small{ \textcolor{softgreen}{+3.45}}& 37.29 \small{ \textcolor{softgreen}{+1.72}} & 64.35 \small{\textcolor{softgreen}{+4.85}} 
\\
\midrule[0.5pt]
\multicolumn{1}{c}{\multirow{6}{*}{Java}}
&  No Rag & 
0.00  & 25.92 & 0.05 & 17.48 &
0.00 & 25.46 & 0.05& 17.61&
0.00 & 25.17 & 0.00 & 17.23

\\
& Shift Rag &
6.45  & 54.84 & 12.11 & 43.75 &
6.08  & 44.73 & 10.27  & 36.46 & 
7.11& 50.96 & 12.72 & 41.68
\\
& Vanilla Rag &
9.68& 55.71& 15.71& 45.71&
10.47& 47.09& 15.29& 39.93 &
11.31 & 51.48& 17.81& 44.09 
\\
&  {\ours} & 
12.16  & 54.16 & 18.37  & 45.47  &
11.50  & 44.90 & 16.22  & 38.95  &
13.32  & 48.04 & 19.54  & 42.34  
\\
\cmidrule(lr){2-14}
&  Repocoder & 
11.22& 56.89& 17.72& 47.41&
10.85& 47.93& 16.18& 41.50 &
13.60 & 52.17& 19.87& 45.14 
\\
&  Repocoder + {\ours} & 
11.50 \small{\textcolor{softgreen}{+0.28}} & 56.09 \small{\textcolor{softred}{-0.80}}& 17.72 \small{\textcolor{softgray}{0.00}} & 46.96 \small{ \textcolor{softred}{-0.45} }&
11.27 \small{\textcolor{softgreen}{+0.42}} & 46.53 \small{\textcolor{softred}{-1.40}}& 16.41 \small{\textcolor{softgreen}{+0.23} }& 40.47 \small{\textcolor{softred}{-1.03}} &
14.26 \small{ \textcolor{softgreen}{+0.66}} & 50.79 \small{\textcolor{softred}{-1.38}}&20.62 \small{ \textcolor{softgreen}{+0.75}} & 44.38 \small{\textcolor{softred}{-0.76}} 
\\
\bottomrule
\end{tabular}
}

\label{tab:Crossfile}
\end{table*}

\subsection{Results and Analysis}
\label{sec:results}

To evaluate the performance of {\ours} on repository-level code completion, we have formulated the following four research questions (RQs):
\begin{itemize}
    \item \textbf{RQ1.Effectiveness:} 
    How does \textsc{\ours} perform compared to other methods?
    
    \item \textbf{RQ2.Cost Analysis:} 
    How does \textsc{\ours}'s resource consumption compare to GraphCoder?
    
    \item \textbf{RQ3.Synergistic Gain Property:} 
    How does \textsc{\ours} perform when integrated orthogonally with other methods that provide cross-file context?
    
    \item \textbf{RQ4.Token Compression:} 
    Does HF\_OP successfully compress the similarity code reference?
\end{itemize}

\subsubsection{For RQ1: {\ours} outperforms other methods in terms of effectiveness.}
As shown in Table~\ref{tab:combined_results}, the evaluation results across both datasets demonstrate the superior performance of {\ours}. On the RepoEval-Updated dataset, {\ours} outperforms the strongest baseline, Graphcoder, in semantic code match (ES) and identifier matching (ID\_EM, ID\_F1), with notable improvements. On the more challenging CrossCodeEval dataset, {\ours} surpasses the other leading baseline (Repocoder), achieving average gains of +1.50 EM, +0.77 ES, +1.11 ID\_EM, and +0.61 ID\_F1. They indicate that {\ours} can more effectively utilize context to generate code with higher semantic accuracy and more appropriate identifiers. For Java code completion on CrossCodeEval, {\ours} achieves better exact match (EM) and identifier exact match (ID\_EM) while showing slightly lower scores in edit similarity (ES) and identifier F1 (ID\_F1). This discrepancy is likely due to the static typing and complex structure of Java, which can lead to logically correct but syntactically variant code. Minor structural deviations (e.g., bracket placement) are thus more penalized by strict text-based metrics like ES and F1. Besides, the smaller gains on RepoEval-Updated compared to CrossCodeEval are due to its higher in-project code similarity, which limits retrieval diversity.

\subsubsection{For RQ2: {\ours} is cost-optimized.}
To evaluate the efficiency of our retrieval-augmented approach under constrained resources, we investigate how the number of retrieved examples (top\_k) affects accuracy using CodeGen25-7B with GraphCoder as the baseline. As shown in Figure~\ref{fig:token_cross}, our method consistently outperforms the baseline at lower top\_k values, particularly under the most constrained setting (top\_k=2-3), where it surpasses GraphCoder across all Python metrics and improves three out of four Java metrics (EM, ES, ID\_EM). 
Performance saturates at top\_k = 3–4, with no benefit from retrieving 10 examples, proving that adding more context is ineffective. This highlights the value of our method, which compresses knowledge to sustain high accuracy using fewer cases, reducing token use without performance loss. Overall, {\ours} enables more efficient and effective code completion, particularly when resources are limited.

\subsubsection{For RQ3: \textsc{\ours} shows synergistic gains when orthogonally integrated with other methods}
We examine two prominent methods that demonstrate exceptional performance in cross-file scenarios. 
\textbf{(1) Repocoder}~\cite{zhang-etal-2023-repocoder}, distinct from the original, assumes that if code snippets are similar, their subsequent content is also likely relevant. In the next search round, it specifically gets the code following those similar snippets (hereafter referred to as Repocoder). \textbf{(2) Draco}~\cite{cheng-etal-2024-dataflow}, analyzes code to create entity dependency graphs, allowing detailed background knowledge retrieval. It then uses this knowledge to create structured prompts. Currently, Draco only works with Python.
As shown in Table~\ref{tab:Crossfile}, adding our method significantly boosts all four Python metrics (by $3.42$ to $4.52$) compared to using Repocoder or Draco alone. For Java, our method improves EM by 0.45 and ID\_EM by $0.33$ over Repocoder, showing {\ours} exhibits significant synergistic gain property with existing cross-file methods. 

\subsubsection{Case Study in Synergistic Gain Property}
To better understand the causes of synergistic gains, we conducted a detailed analysis of our experimental cases. We found that the performance boost is not merely a general improvement but a direct result of how {\ours} strategically manages the context window.   Specifically, {\ours} utilizes its HF\_OP to compress and filter redundant knowledge; this reduction in ``noise” is what creates the necessary space for Draco’s detailed data-flow graphs or RepoCoder’s retrieved snippets to fit into the context window without being truncated.  In the ``Draco + {\ours}” configuration, while Draco provides deep semantic modeling, it often struggles when code intent is ambiguous.   {\ours} resolves this by injecting semantically aligned examples that serve as ``intent hints.”   Because {\ours} efficiently resolves the ``type inference chain breakage” with concise symbolic relationships, the model can ingest Draco’s complex control-flow graphs in their entirety, allowing the two modules to compensate for each other's weaknesses.  Similarly, the ``RepoCoder + {\ours}” combination succeeds because {\ours} acts as a filter for external adaptability.   While RepoCoder often introduces knowledge that clashes with local APIs, {\ours}’s disambiguation module clarifies identifier dependencies beforehand.   By compressing these dependencies into a streamlined format, {\ours} ensures that RepoCoder’s most relevant coding patterns are prioritized within the prompt window rather than being cut off by irrelevant external noise.   This precise causal link—where {\ours}’s compression of A (symbolic noise) enables B (critical external context) to remain intact—is the primary driver of the observed synergistic performance.

\begin{table}[h]
\centering
\caption {Quantitative analysis of top\_k and token when max\_num\_tokens is 2048.}
\resizebox{\linewidth}{!}{
\begin{tabular}{lcccccc}
\toprule
\multicolumn{1}{c}{\multirow{2}{*}{\textbf{Method}}}&
\multicolumn{2}{c}{\textbf{Codegen2-7b}} & \multicolumn{2}{c}{\textbf{deepseek-coder-6.7b-instruct}} 
& \multicolumn{2}{c}{\textbf{CodeLlama-7b-Instruct}}
\\
\cmidrule(lr){2-3} \cmidrule(lr){4-5} \cmidrule(lr){6-7}
&  \#In &  \#Out & \#In & \#Out &   \#In &  \#Out 
\\
\midrule[0.5pt]
No RAG 
& 1397.38 & 50.44 & 1383.12 & 95.22 &1388.38 &50.146
\\
Shifted RAG
& 821.627 & 95.15 & 778.52 & 93.48& 763.93 &96.86 
\\
HF\_OP (top\_k=10)
& 1665.50 & 88.15 & 1621.92 & 90.86 &1611.37 & 96.99 
\\
HF\_OP (top\_k=4)
&  \textbf{1644.26} & 89.20 & \textbf{1596.83} & 88.57 & \textbf{1590.55} & 98.06 
\\
\bottomrule
\end{tabular}
}
\label{tab:token}
\end{table}

\begin{table}[ht]
\centering
\caption{Hit@k Performance on RepoEval\_Updated and CrossCodeEval Datasets. ``No Comp.” denotes the original retriever without any compression processing.}
\label{tab:hit_combined}
\resizebox{\linewidth}{!}{
\begin{tabular}{ccccccc}
\toprule
\multicolumn{1}{c}{\textbf{Dataset}} & 
\multicolumn{1}{c}{\textbf{Language}} & 
\multicolumn{1}{c}{\textbf{Method}} & 
\multicolumn{1}{c}{\textbf{hit@1}} & 
\multicolumn{1}{c}{\textbf{hit@5}} & 
\multicolumn{1}{c}{\textbf{hit@10}} \\
\midrule[0.5pt]
\multirow{4}{*}{RepoEval\_Updated}
& \multirow{2}{*}{python} 
& No Comp. & 25.40\% & 32.15\% & 34.35\% \\
& & HF\_OP & 25.00\% & 34.50\% & 36.45\% \\
\cmidrule(lr){2-6}
& \multirow{2}{*}{java} 
& No Comp.& 12.69\% & 19.81\% & 22.25\% \\
& & HF\_OP & 15.12\% & 21.94\% & 23.19\% \\
\midrule[0.5pt]
\multirow{4}{*}{CrossCodeEval}
& \multirow{2}{*}{python} 
& No Comp. & 0.11\% & 0.26\% & 0.38\% \\
& & HF\_OP & 2.61\% & 3.97\% & 4.16\% \\
\cmidrule(lr){2-6}
& \multirow{2}{*}{java} 
& No Comp. & 0.70\% & 1.36\% & 1.36\% \\
& & HF\_OP & 0.84\% & 1.36\% & 1.54\% \\
\bottomrule
\end{tabular}
}
\end{table}

\subsubsection{For RQ4: HF\_OP successfully compresses code reference.}
To quantitatively validate the effectiveness of our HF\_OP method in compressing reference code knowledge, we analyze the retrieval performance alongside the token efficiency. As table~\ref{tab:token} shown, setting \textit{top\_k} to 4 instead of 10 reduces input token consumption by approximately 22.38 tokens per task on average without compromising output quality.
Crucially, this token saving is achieved while maintaining or even enhancing retrieval accuracy. The Hit@k results on both the RepoEval-Updated and CrossCodeEval datasets confirm this. As shown in Table~\ref{tab:hit_combined}, HF\_OP method consistently matches or surpasses the baseline No Comp. across most metrics. On the \textbf{RepoEval-Updated} dataset, HF\_OP improves the hit@5 and hit@10 scores for Python and for Java across all three metrics. More substantial gains are observed on the \textbf{CrossCodeEval} dataset, where HF\_OP delivers a dramatic improvement, especially in Python. These results demonstrate that by filtering redundant and homogeneous cases, HF\_OP not only frees up token budget but also increases the knowledge density of the retained reference snippets, leading to a more effective retrieval set. 

\subsection{Ablation Study}
\begin{figure}
\centering
\includegraphics[width=\linewidth]{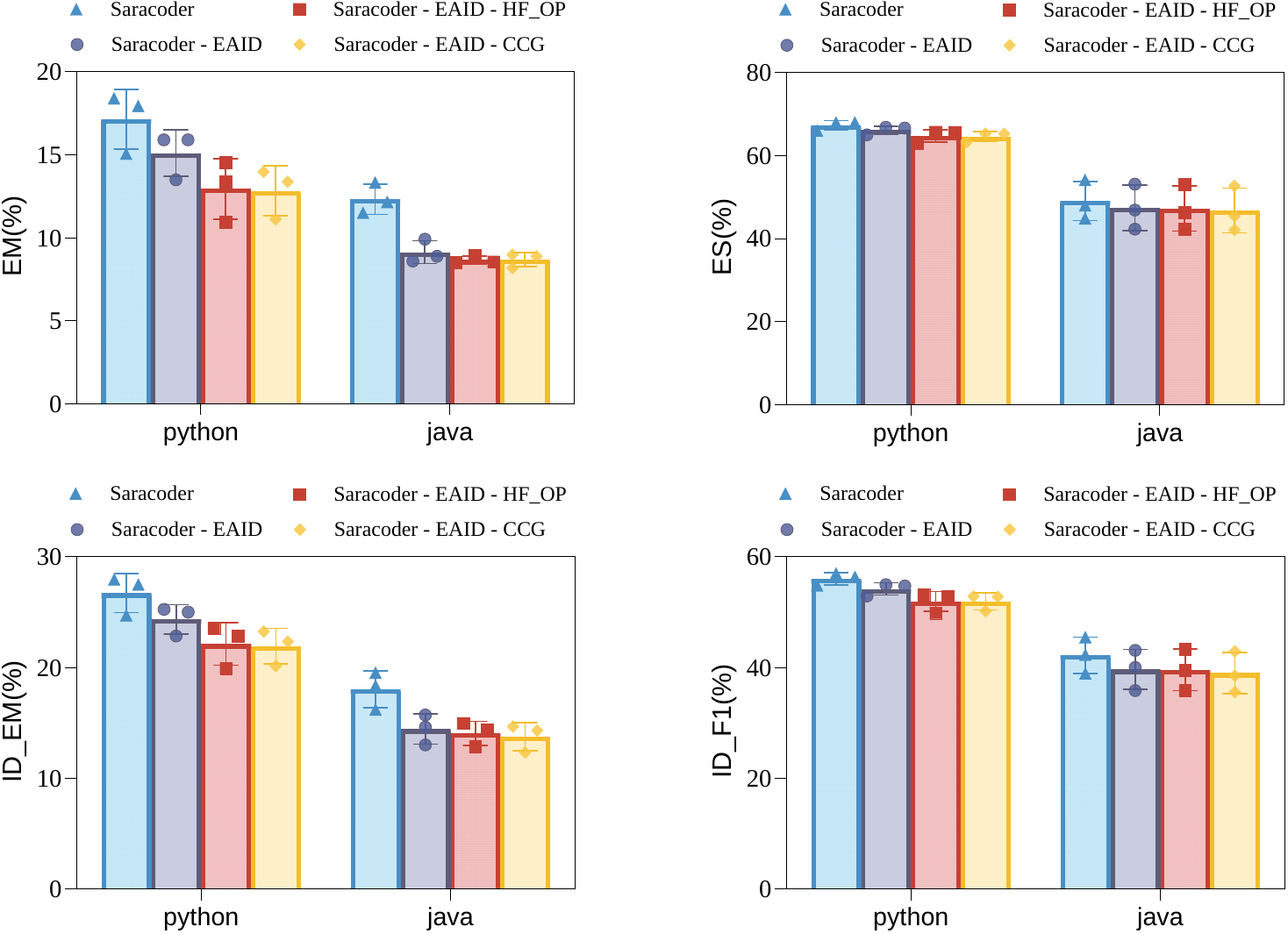}
\caption{Ablation study. (Each three-data-point group represents CodeGen2-7B, CodeGen2.5-7B, and CodeLlama-7B-Instruct models. Bar lengths show their average performance, with I-shaped error bars indicating standard deviation) 
}
\label{fig:ablation}
\end{figure}

To understand the importance of each part of {\ours}, we conduct ablation tests on the CrossCodeEval dataset (Figure~\ref{fig:ablation}). ``-EAID” indicates disabling  \textbf{External-Aware Identifier Disambiguator}, resulting in the loss of external dependency integration capabilities; ``-HF\_OP” denotes removing \textbf{Hierarchical Feature Optimization}, canceling the similar fragment screening mechanism; ``-CCG” indicates disabling \textbf{the code context graph}, so it lost the understanding of code structure. The ablation experiments demonstrate that the complete {\ours} achieves optimal performance, with all components positively contributing to repository-level completion. 

\section{Conclusion}

In this paper, we presented {\ours}, a resource-optimized framework for repository-level code completion. We addressed a key challenge where limited context windows force a trade-off between semantic reference snippets and cross-file dependency definitions, a conflict often exacerbated by redundant retrieval in existing methods. To resolve this, {\ours} employs a Hierarchical Feature Optimization to compress reference code into high-density summaries and integrates an External-Aware Identifier Disambiguator to inject precise dependency knowledge. This approach maximizes the synergy between both contextual clues, significantly improving completion performance. While currently validated on Java and Python, our future work will focus on expanding language coverage and exploring cross-language completion.

 


\bibliographystyle{IEEEtranN}
\bibliography{reference}

\vspace{-1cm}
\begin{IEEEbiography}[{\includegraphics[width=1in,height=1.25in,clip,keepaspectratio]{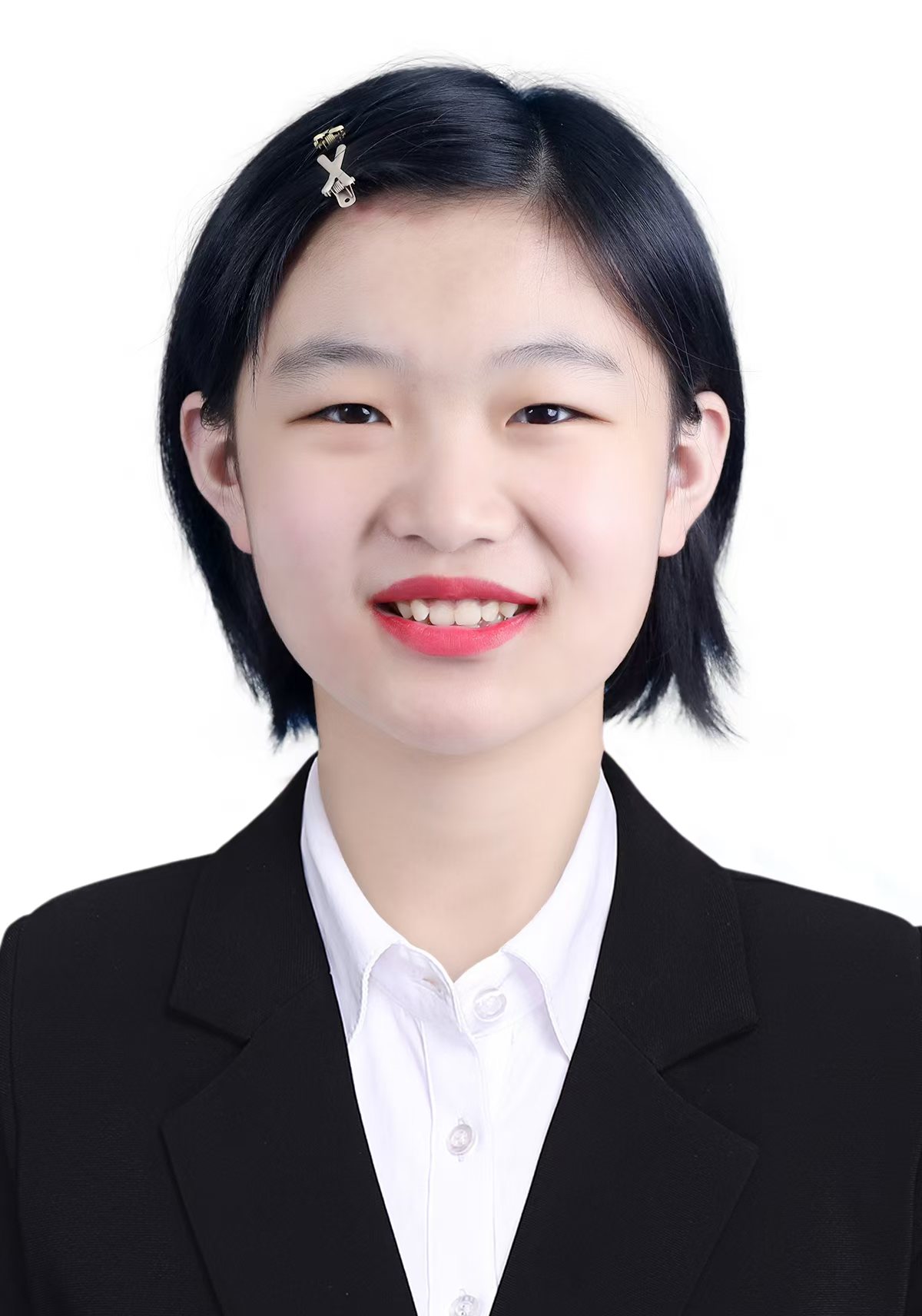}}]{Xiao-Han Chen} graduated from the College of Computer Science and Technology, Nanjing University of Aeronautics and Astronautics, with a Bachelor's degree. She is currently pursuing a Master's degree at the Institute of Information Engineering, Chinese Academy of Sciences, in the State Key Laboratory of Cyberspace Security Defense. Her primary research areas are retrieval-augmented generation (RAG) and database security and protection.
\end{IEEEbiography}

\vspace{-1cm}
\begin{IEEEbiography}[{\includegraphics[width=1in,height=1.25in,clip,keepaspectratio]{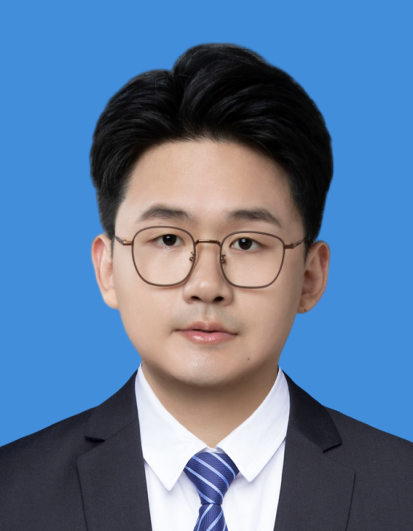}}]{Xiang Chen} received the PhD degree from the College of Computer Science and Technology at Zhejiang University in 2024. He is currently a professor in the College of Computer Science and Technology at Nanjing University of Aeronautics and Astronautics, leading a research group focused on multimodal and language intelligence technologies. He has served as a PC/AC for several prestigious conferences, including ICLR, NeurIPS, ACM MM, ACL ARR, et al. His current research interests include LLMs and knowledge graphs.
\end{IEEEbiography}

\vspace{-1cm}
\begin{IEEEbiography}[{\includegraphics[width=1in,height=1.25in,clip,keepaspectratio]{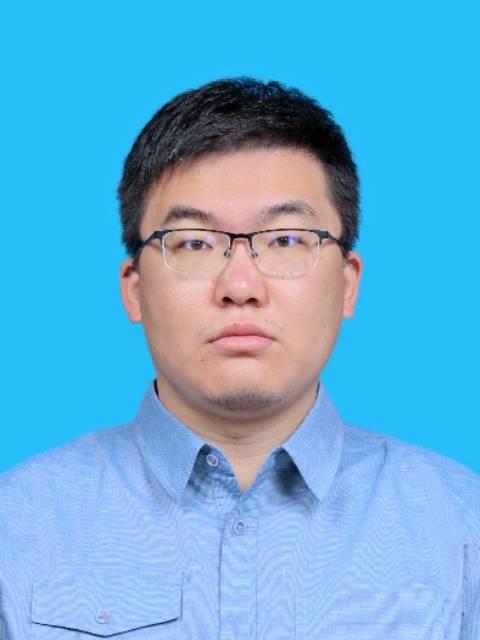}}]{Yu Tian} graduated from the University of Chinese Academy of Sciences in 2023 with a Ph.D. degree. His primary research areas are multimodal large language models and time series prediction.
\end{IEEEbiography}

\vspace{-1cm}
\begin{IEEEbiography}[{\includegraphics[width=1in,height=1.25in,clip,keepaspectratio]{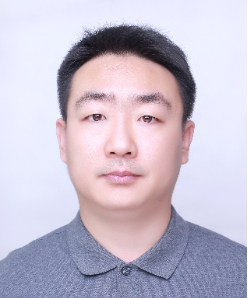}}]{Piji Li} received the BEng and MEng degrees from Shandong University in 2009 and 2012, respectively, and the PhD degree from the Chinese University of Hong Kong in 2018. He is currently a professor in the College of Computer Science and Technology at Nanjing University of Aeronautics and Astronautics and lead the NUAA NLP Group. Previously, He was a senior researcher at Tencent AI Lab. His research interests are mainly in LLMs and dialogue systems.
\end{IEEEbiography}

\vspace{-1cm}
\begin{IEEEbiography}[{\includegraphics[width=1in,height=1.25in,clip,keepaspectratio]{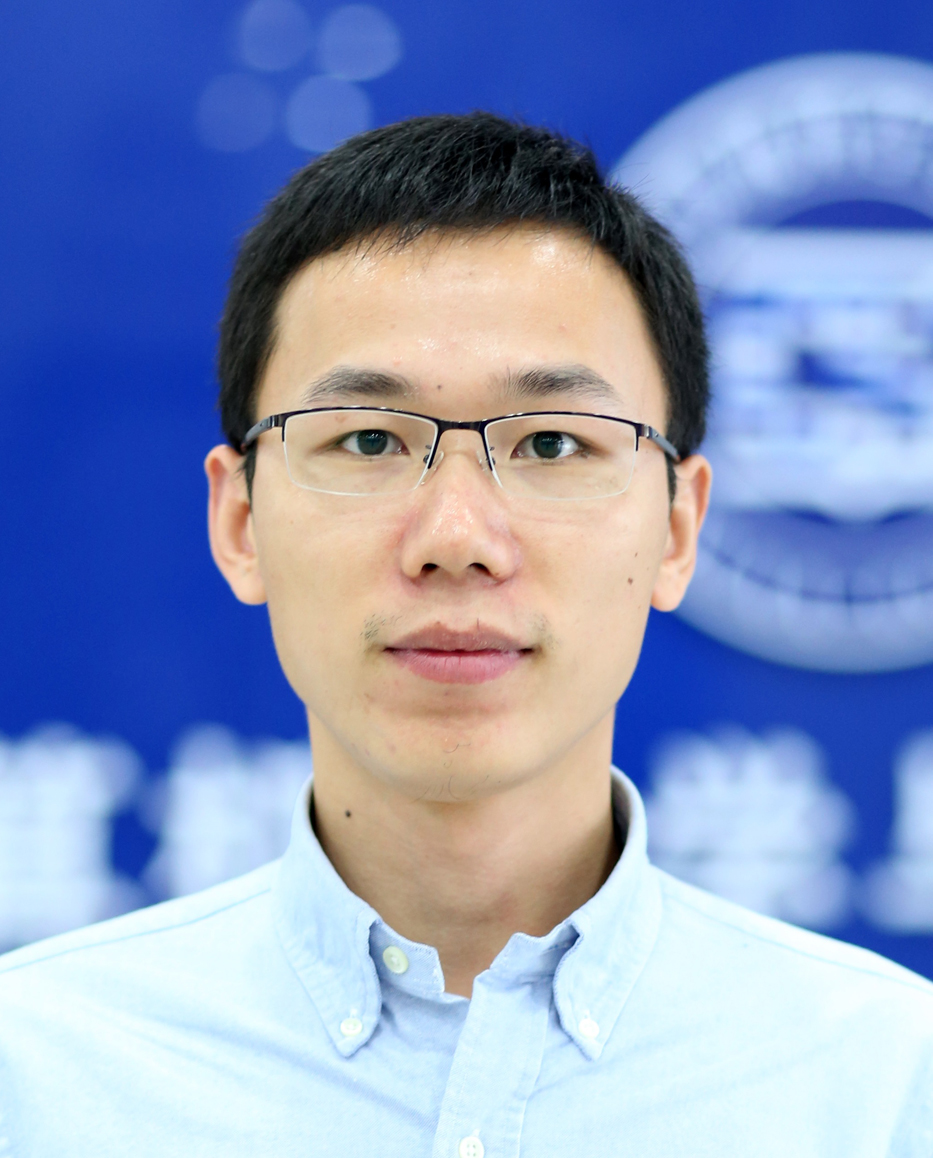}}]{Sheng-Jun Huang} received the BEng and PhD degrees in computer science from Nanjing University in 2008 and 2014, respectively, under the supervision of Prof. Zhi-Hua Zhou. He is currently a professor at Nanjing University of Aeronautics and Astronautics, the dean of College of Computer Science and Technology. His research interests include machine learning and data mining.
\end{IEEEbiography}

\vspace{-15cm}
\begin{IEEEbiography}[{\includegraphics[width=1in,height=1.25in,clip,keepaspectratio]{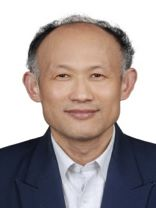}}]{Songcan Chen} received the PhD degree from Nanjing University of Aeronautics and Astronautics in 1997. He is currently the Director of the Machine Learning Committee of the Chinese Artificial Intelligence Association, the Executive Director of the Chinese Artificial Intelligence Association, and the Executive Vice President of the Jiangsu Artificial Intelligence Association. He is also a Fellow of the International Association for Pattern Recognition (IAPR Fellow) and a Fellow of the Chinese Artificial Intelligence Association (CAAI Fellow). He has published more than 170 SCIE papers in international mainstream academic journals including IEEE Transactions. According to Google Scholar statistics, his papers have been cited more than 20,000 times, with an H-index of 66. He has been selected into the Elsevier China Highly Cited Scholars List for 10 consecutive years from 2014 to 2024. His current research interests include machine learning and data mining.
\end{IEEEbiography}

\end{document}